\documentclass[doublecol]{epl2}
\usepackage{epsfig,color}
\usepackage{xcolor}

\usepackage{ulem}

\title{Dipolar needles in the microcanonical ensemble: evidence of spontaneous magnetization
and ergodicity breaking}

\author{George Miloshevich\inst{1,2} \and Thierry Dauxois\inst{3} \and Ramaz Khomeriki\inst{1,4}
\and Stefano Ruffo\inst{3,5}}
\shortauthor{G. Miloshevich \etal}

\institute{
  \inst{1} Department of Physics, Faculty of Exact
and Natural Sciences, Tbilisi State University, 0128 Tbilisi,
Georgia\\
  \inst{2} Department of Physics, The University of Texas at
Austin, Austin TX 78712, USA \\
\inst{3} Laboratoire de Physique de l'ENS Lyon
, Universit\'e de Lyon, CNRS, 46, all\'ee d'Italie, 69007 Lyon,
France\\
\inst{4} Max-Planck Institute for the Physics of Complex Systems,
N\"othnitzer Str. 38, 01187 Dresden, Germany \\
\inst{5} Dipartimento di Fisica e Astronomia and CSDC,
Universit\`a di Firenze, CNISM and INFN, via G. Sansone, 1, Sesto
Fiorentino, Italy}


\abstract{We have studied needle shaped three-dimensional classical spin systems
with purely dipolar interactions in the
microcanonical ensemble, using both numerical simulations and
analytical approximations. We have observed spontaneous
magnetization for different finite cubic lattices. The transition
from the paramagnetic to the ferromagnetic phase is shown to be first-order.
For two lattice types we have observed magnetization flips
in the phase transition region. In some cases, gaps in the accessible values of magnetization
appear, a signature of the ergodicity breaking found for systems with long-range
interactions. We analytically explain these effects
by performing a nontrivial mapping of the model Hamiltonian onto a
one-dimensional Ising model with competing antiferromagnetic
nearest-neighbor and ferromagnetic mean-field interactions.
These results hint at performing experiments on isolated dipolar needles
in order to verify some of the exotic properties of systems with long-range
interactions in the microcanonical ensemble.}

\begin{document}

\maketitle

Systems with long-range interactions, such as gravitational,
Coulomb and magnetic systems, are of fundamental and practical
interest because of their exotic statistical properties including
ensemble inequivalence, negative specific heat, temperature jumps,
ergodicity breaking, etc.~\cite{rep}  Recently, a number of
mean-field type models have been developed which are very
convenient for analytical understanding~\cite{BEGmodel,mukruf}.
However, up to now, the connection to real physical systems has
not been seriously addressed (see, however, Refs.~\cite{JulienBarre,campa,DuccioCatalioti}
for some progress in this direction).
It is therefore crucial to propose experimentally testable effects.

Dipolar force is one of the best candidates for experimental and
theoretical studies of long-range interactions~\cite{long}. For
instance, experimental studies have been performed on layered spin
structures~\cite{sievers}. For these systems, intralayer exchange
is much larger than the interlayer one: hence, every layer can be
identified as a single macroscopic spin. As a consequence, dipolar
forces between layers are dominant and one
can describe the system with an effective long-range
one-dimensional model~\cite{campa}. However, in order to perform a
careful study of the statistical properties of such samples, one
should simulate all the spins in each layer, which is computationally heavy.

Alternatively, one can consider purely dipolar systems known as
dipolar ferromagnets \cite{ros,Corruccini}, where dipolar effects prevail
over short-range exchange interactions. Long-range dipolar orientational
order is also found theoretically for dipolar fluids confined in ellipsoidal
geometries \cite{Dietrich}.
More recently, dipolar ferromagnetism has also been measured at ambient temperature
in assemblies of closely-spaced cobalt nanoparticles \cite{ref}.

It has been pointed out long ago \cite{lut} that body centered
cubic (bcc) or face centered cubic (fcc) needle like samples
should display spontaneous magnetization, while simple cubic (sc) lattices
can be ordered only antiferromagnetically. On the
other hand, it was later argued that dipolar systems cannot show
nonzero magnetization in the thermodynamic limit \cite{grif}, i.e. as
a bulk property. All these theoretical studies were performed within the
canonical ensemble, but we know that ensemble inequivalence is
expected to be present also for dipolar systems~\cite{rep}.
This means that the phase diagram of dipolar systems can be different in the
microcanonical ensemble: the location of phase transition points can vary, temperature
jumps may appear and ergodicity may be broken~\cite{celardo,mukruf,bouchet}. It is therefore important to
perform a study on samples with needle shape in the microcanonical ensemble.
Experimentally, microcanonical ensemble measurements imply the realization of an
isolated sample, or looking at time-scales that are fast with respect to the
energy exchange rate with environment.

In this Letter, we study the microcanonical dynamics of dipolar
needles via numerical simulations and analytical approximations. We want to check whether
such systems can display spontaneous magnetization and study the nature
of the paramagnetic/ferromagnetic phase transition.

Systems of classical spins with only dipolar interactions are described by the following Hamiltonian
\begin{equation}
{\cal H}=\frac{\varepsilon}{2}\sum\limits_{i\neq
j}\frac{a^3}{r_{ij}^3}\left(\vec S_{i}\cdot\vec S_{j}
-3\frac{(\vec S_{i}\cdot\vec r_{ij})(\vec S_{j}\cdot\vec
r_{ij})}{r_{ij}^2}\right), \label{11}
\end{equation}
where $|\vec S_i|=1$ is a unit vector located at the $i$-th
lattice site, $\vec r_{ij}$ is the displacement vector between
$i$-th and $j$-th site, $a$ stands for the lattice spacing and
$\varepsilon=\mu_0\sigma^2/(4\pi a^3)$ is an energy scale for dipolar
interactions: for instance, in the case of cobalt nanoparticles with
magnetic moment $\sigma\sim 2 \cdot 10^5 \mu_B$ and separation length
$a\sim 20nm$ \cite{ref}, this energy could be as large as $2500$K
($\mu_0$ is vacuum permeability and $\mu_B$ Bohr magneton).

The time evolution of the unit vector $\vec S_i$ is described by the torque equation
\begin{equation}
\frac{d {\vec S_i}}{dt}=\gamma{\vec S_i} \times {\vec H}_i \qquad
\mbox{where}\qquad{\vec H}_i=-\frac{1}{\sigma}\frac{\partial{\cal
H}}{\partial{\vec S_i}}. \label{torque1}
\end{equation}
Here, ${\vec H}_i$ is the local magnetic field acting on the spin
attached to the $i$-th lattice site, $\gamma$ is the particle
gyromagnetic ratio and, in numerical simulations, we measure
time in units of $\mu/(\gamma \varepsilon)$.

\begin{figure}[t]
\epsfig{file=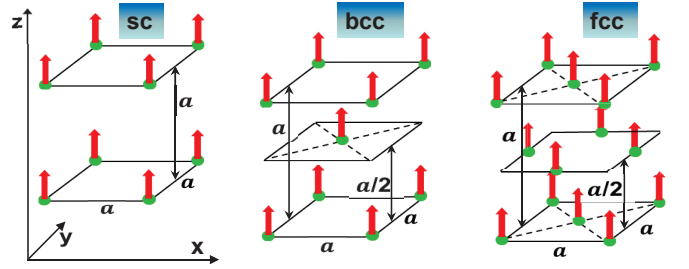,width=1\linewidth} \vskip -0.3truecm
\caption{Cubic lattices considered in numerical simulations:
simple cubic (sc), body centered (bcc) and face centered
(fcc) cubic, respectively. The arrows indicate the
initial direction of the spins while $a$ is the lattice spacing.}
\label{lattice}
\end{figure}

In numerical experiments, we solve the torque equation for spins
on sc, bcc and fcc lattices shown in Fig.~\ref{lattice}.
Initially, the spins are aligned along the main axis of the sample,
as shown in the figure. In the course of time, we monitor the
three components of the average magnetization $\vec m=(1/N)\sum_{i=1}^N\vec S_i$,
where $N$ is the number of spins over which we perform an average.

While for numerical simulations we directly use
Hamiltonian~(\ref{11}) with the torque equation, our analytical
approach is based on heuristic approximations by which we  are
able to map the main properties of Hamiltonian~(\ref{11}) onto those
of the simple one-dimensional mean-field model studied in
Ref.~\cite{mukruf}.

We consider samples elongated in the $z$-direction. This is the
ordering direction of the spin system because the demagnetizing
field is smaller along this axis. We follow the standard treatment in
Refs.~\cite{landau,book} by dividing the sums in Hamiltonian~(\ref{11}) in
two parts: The first part is the sum restricted only
to a neighborhood of a site in the same layer (the reason this is done will be clear below), the second part is the sum over the
remaining portion of the sample. We treat the latter sum,
which takes into account the long-range character of the dipolar
interaction, via a continuum approximation~\cite{book}, which gives
\begin{equation}
{\cal H}_{cont}=-\frac{\varepsilon
a^3}{2\sigma^2}\int\left[\frac{4\pi}{3}\left(\vec
M(\vec r)\right)^2+\vec H^m(\vec r)\vec M(\vec r)\right]d^3r,
\label{cont}
\end{equation}
where the magnetization density $\vec M (\vec r)$ is obtained by a local average over
a macroscopic number of spins and $\vec H^m(\vec r)$ is the demagnetizing field. In the case of ellipsoidal
samples, there exists a uniform solution with demagnetizing field proportional to magnetization density
$\vec H^m=-4\pi\hat C \vec M$, where
\begin{equation}
\vec M=\frac{\sigma}{V}\sum_{i=1}^N\vec S_i
\label{mag}
\end{equation}
and $\hat C$ stands for the demagnetization tensor.
For simplicity, we neglect the transversal components of the spin vectors, only the longitudinal
components~$S_i^z$ are considered. As a further simplification, we
assume that the longitudinal component takes only two values
$S_i^z=\pm 1$, i.e. we reduce to Ising spins. After making such a crucial
simplification, only the magnetization density along the $z$-axis is nonzero and we can
easily perform the integral in formula~(\ref{cont}), obtaining
\begin{equation}
{\cal H}_{cont}=-\frac{\varepsilon
a^3}{2Nv_0}\left[\frac{4\pi}{3}-4\pi
C_z\right]\biggl(\sum\limits_{i=1}^NS^z_{i}\biggr)^2,
\label{cont1}
\end{equation}
where we have substituted $V=Nv_0$, $v_0$ being the volume per spin. It is important to point out that what we mean by volume is nontrivial in case of finite systems. We define the volume as the box that encloses the crystal so that the size of the box is obtained by matching total energies of the effective and discrete models. It turns out that in all lattices considered below this requires the box to be outstretching by a half a lattice constant from the spins at the boundary. This also has an effect on the definition of the aspect ratio.

\begin{figure}[t]
\epsfig{file=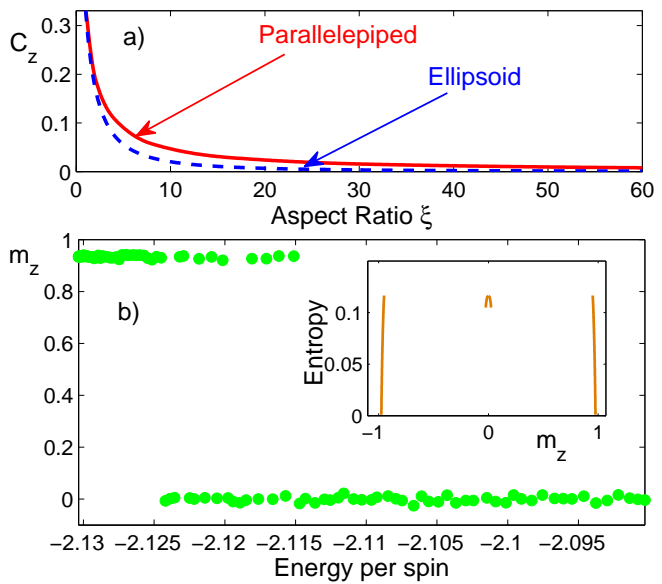,width=1\linewidth} \vskip -0.3truecm
\caption{a) Demagnetizing coefficient~$C_z$ computed numerically
for a parallelepiped using formula~(\ref{Nz1}) (solid line) versus
the aspect ratio of the needle. We also plot the exact curve for
an ellipsoid (dashed line), whose expression is given in
formula~(\ref{formulaelliposid}). b) Final values  of the
$z$-component of the magnetization~$m_z$ versus the energy per spin
(in units of $\varepsilon$), obtained in numerical simulations of the
sc lattice in Fig.~\ref{lattice}. In the inset, we plot entropy
versus magnetization of the effective Hamiltonian~(\ref{3}) at a
typical energy in the range of the transition.}
\label{NN}
\end{figure}

As far as we consider only the $z$-components of the spins, it is easy to
see that, for all considered lattices, in each layer transversal
to the $z$-component of the sample, the coupling among the spins
is antiferromagnetic. On the contrary, the coupling between
neighboring spins in close transversal layers is ferromagnetic.
This latter contribution is included in the ferromagnetic type
coupling (\ref{cont1}). Thus, incorporating nearest neighbor
coupling terms and the terms appearing from the continuum
approximation (\ref{cont1}), we can reduce Hamiltonian~(\ref{11})
to the following effective Hamiltonian
\begin{equation}
{\cal H}_{eff}=-\frac{K}{2}\sum\limits_{\langle i,
i'\rangle}\left(S^z_{i}S^z_{i'}-1\right)
-\frac{J}{2N}\biggl(\sum\limits_{i=1}^NS^z_{i}\biggr)^2, \label{3}
\end{equation}
where the lattice dependent short-range coupling $K$ is heuristically estimated below,
$\langle i,i'\rangle$ means that the sum is restricted only
to nearest-neighbors in the transversal layers and the
ferromagnetic mean-field coupling constant $J$ is given by the
following expression
\begin{equation} J=\frac{4\pi\varepsilon a^3 (1-3C_z)}{3v_0},
\label{J}
\end{equation}
while the demagnetizing coefficient $C_z$ is given by the following
integral~\cite{book}
\begin{equation}
C_z=-\frac{1}{4\pi V}\int_Vd^3r \int_V
d^3r_1\frac{\partial^2}{\partial z^2}\left(\frac{1}{|\vec r-\vec
r_1|}\right). \label{Nz1}
\end{equation}
The demagnetizing coefficient~$C_z$ is $1/3$ if the sample length
$L$ coincides with the lattice spacing~$a$, while it tends to zero
if the aspect ratio $\xi=(L+a)/(2a)$ tends to infinity. In
Fig.~\ref{NN}a, we plot the dependence of this coefficient on the
aspect ratio, comparing the case of a parallelepiped with that of
an ellipsoid, for which the exact expression is
\begin{equation}
C_z^{ell}=\frac{1-b^2}{2b^3}\left(\ln \frac{1+b}{1-b}-2b\right) ,
\label{formulaelliposid}
\end{equation}
where $b=\sqrt{1-1/\xi^2}$. In all the
estimates below, we use the demagnetizing coefficients of the parallelepiped,
which gives a better quantitative agreement between analytical results
derived from (\ref{3}) and numerical simulations performed on (\ref{11}).

Let us remark that the second term in the effective
Hamiltonian~(\ref{3}) contains the $z$-component of the average
magnetization  $m_z=(1/N)\sum_{i=1}^NS_i^z$, and its typical size
can be varied by changing the aspect ratio~$\xi$, i.e. the
length~$L$ of the sample.

We emphasize that the effective Hamiltonian~(\ref{3}) is the same
as that in formula (1) of Ref.~\cite{mukruf}. In the following, we
will use results obtained for this Hamiltonian to discuss the
phase diagram of model~(\ref{11}). Our approach consists in
performing an estimate of the values of the couplings $K$ and $J$
based on features of the finite sample. In Ref.~\cite{mukruf}, it
is proven that Hamiltonians of type~(\ref{3}) undergo a phase
transition of the ferromagnetic type. This phase transition is
of second order if both couplings~$K$ and~$J$ are positive. It
becomes first order if the coupling~$K$ is sufficiently negative,
which favors locally the antiferromagnetic phase. The phase
transition is present for values above $K/J=-0.5$, while for
values below the system is always in the paramagnetic phase.
Hence, what determines the presence of the phase transition in
model~(\ref{11}) is the ratio $K/J$. This ratio can be estimated
using the above expression of $J$ for particular choices of the
lattice (e.g. those of Fig.~\ref{lattice}) and by a rough estimate
of the coupling constant~$K$.

For {\it simple cubic lattices} (see Fig.~\ref{lattice})
which have four spins in the transversal layer, the coupling
constant $K=-2\varepsilon$. For $\xi\rightarrow \infty$,
Eq.~(\ref{Nz1}) leads to $J=4\pi\varepsilon/3$, since the volume per
spin in a sc lattice is $v_0=a^3$. Thus, $K/J=-3/(2\pi)>-0.5$
and, therefore, we can expect the presence of a ferromagnetic
phase.

In order to verify this prediction, we have performed numerical
simulations of model (\ref{11}), on a $2\times2\times50$ sc
lattice, starting from a fully magnetized initial state, i.e. all
spins pointing strictly along the $z$-axis.
We then vary the energy of the initial state by adding random
transversal components to the spins. We let the system relax
towards a stationary state (this typically happens at times of the
order $10^4$) and monitor  the final value of $m_z$ for each value
of the energy. We checked that the final state does not contain
domains by looking at the spatial patterns of the individual
spins. Collecting all these final values of the magnetization, we
plot them as a function of energy per spin in Fig.~\ref{NN}b. We
clearly observe a jump in magnetization from a positive value to
zero and the presence of a coexistence region, where both a
paramagnetic and a ferromagnetic phase are present, in full
accordance with the predictions of the effective Hamiltonian
(\ref{3}) that the phase transition is first-order in the
microcanonical ensemble. For symmetry reasons, we would have observed also the
negative magnetization state if we had prepared the sample with
the spins aligned opposite to the $z$-axis. The existence of this
symmetry is confirmed by looking at the entropy of the effective
Hamiltonian~(\ref{3}) (see Eq.~(3) in \cite{mukruf}) as a function
of~$m_z$, shown in the inset of Fig.~\ref{NN}b. The gaps in the
accessible values of magnetization are a signature of {\it ergodicity
breaking}~\cite{mukruf,celardo,bouchet} . The first order phase transition takes place when the maxima
of the entropy of the paramagnetic and ferromagnetic states are at
the same height.

As we vary the size and the shape of the sample, the couplings $K$
and $J$ in Hamiltonian~(\ref{11}) change. We have then to check
whether we are still in a region of parameters where the phase
transition is present. It happens that for a sc lattice, an
increase of the base size cancels the phase transition and the
system always remains in the paramagnetic phase. Indeed for a
wider base, each spin has four neighbors and thus the effective
antiferromagnetic coupling constant $K=-4\varepsilon$, while the
ferromagnetic mean-field constant remains $J=4\pi\varepsilon/3$ even
for large aspect ratios. Therefore the ratio $K/J\approx -1$ does
not allow for magnetized states to exist. This has been verified
in numerical simulations which show that, even in the case of a
$3\times 3$ base, there is no phase transition.

In order to verify whether other types of lattices can support phase
transitions as suggested in \cite{lut}, we have performed
simulations for the bcc and fcc lattices shown in Fig.~\ref{lattice}.
We have shown that these lattices do have a phase transition if
the aspect ratio is large enough, and therefore these dipolar
samples display spontaneous magnetization.

For {\it body centered cubic lattices}, we have four spins in a
layer and one spin in the neighboring layer. In the layer with
four spins, the situation is exactly the same as that of the sc
lattices, while in the layer with one spin there is no intralayer
interaction. Thus, 4/5 of all spins can be treated as in sc
lattices, while one of the five cannot have an antiferromagnetic
coupling. These latter spins form a vertical chain, and their
contribution to the Hamiltonian is a simple energy shift. Thus,
the effective Hamiltonian for this lattice takes the form
\begin{equation}
{\cal H}_{eff}^{bcc}=\frac{4}{5} {\cal H}_{eff}
+\frac{1}{5}NE_0.\label{33}
\end{equation}
The antiferromagnetic coupling constant $K$ is unchanged. On the
contrary, the mean-field ferromagnetic coupling constant $J$
changes since the average volume per spin is now $v_0=4a^3/5$,
which implies that for large aspect ratios $J=5\pi\varepsilon/3$.
As a consequence, $K/J\approx- 0.4$ and, from what we know of Hamiltonian~(\ref{3}),
we can therefore expect many different regimes, contrary to the case of
sc lattices, where the ratio $K/J$ is close to $-0.5$.

In the following, we will need an estimate of the value of the
energy shift $E_0$. Two ferromagnetic contributions appear in this
quantity: the first one comes from the sum over all the sample
while the second one derives from the sum over the neighboring
spins along the vertical chain. For large aspect ratios, one has
approximately $E_0=\left[-5\pi/3-4\right]\varepsilon/2$.

\begin{figure}[t]
\epsfig{file=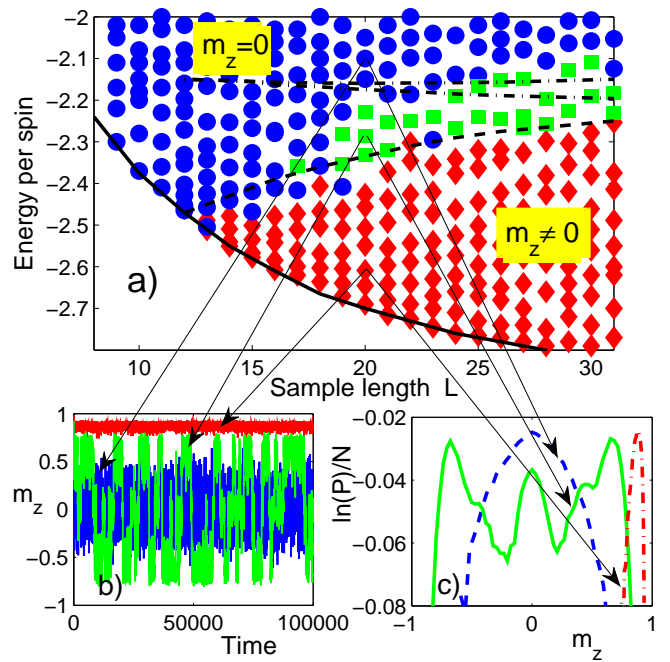,width=1.\linewidth} \vskip -0.3truecm
 \caption{a) Phase diagram of Hamiltonian~(\ref{11}) for the bcc lattice in
Fig.~\ref{lattice}. Circles (blue), diamonds (red)  and squares
(green) represent paramagnetic, ferromagnetic and flipping states,
respectively. The solid line is the minimal energy computed using
Hamiltonian~(\ref{33}), the dashed one is the phase transition
line and the dash-dotted lines are the bounds of the region with
magnetization flips, again computed with the same effective
Hamiltonian. b) Time evolution of the magnetization in different
energy regions as shown by the arrows: $L=20$ with an energy per
spin $-2.6$ (ferromagnetic ), $-2.3$ (flips) and $-2.1$
(paramagnetic); in all graphs energy is scaled in units of
$\varepsilon$. c) Entropy per spin for the three phases in panel b).
} \label{phase}
\end{figure}

We have performed numerical simulations for the bcc lattice of the
full dipolar Hamiltonian~(\ref{11}). We have first concentrated
our attention on detecting the presence of a phase transition. As
control parameters, we use the energy per spin and the length $L$
of the sample. In Fig.~\ref{phase}a, we plot with circles the
paramagnetic states and with diamonds the ferromagnetic ones. The
minimal energy for each sample length is calculable from
Hamiltonian~(\ref{33}) and is shown by the solid line in
Fig.~\ref{phase}a: the agreement with numerical simulations of
Hamiltonian~(\ref{11}) is also very good.

The effective Hamiltonian~(\ref{33}) predicts the location
of the phase transition energies for each value of $K/J>-0.5$ (dashed
line in Fig.~\ref{phase}a). The calculation can be performed by looking at
entropy versus magnetization: near the phase transition
point this curve is characterized by three disconnected humps, one at zero magnetization and two
others at negative and positive magnetizations. On increasing energy
from low values, the entropy at nonzero magnetization increases, while the zero
magnetization entropy decreases: they become equal at the phase transition 
point. One should emphasize that, both below and above the phase transition energy, the 
entropy/magnetization curve is disconnected, showing {\it ergodicity breaking}. 
Only after a further increase of energy, the curve becomes connected with still three
humps: this energy regime corresponds to the region where magnetization $m_z$ flips 
among positive, negative and zero values. This region is delimited by the dash-dotted 
lines in Fig.~\ref{phase}a.

Numerical simulations confirm the presence of a region of
magnetization flips as shown in Fig.~\ref{phase}b, but not
precisely at the same location in the parameter space that we predict. In
Fig.~\ref{phase}c, we plot for three different energies the
entropy per spin $s=(\ln P(m_z))/N$ where $P(m_z)$ is obtained
from the histogram of the magnetization. The first energy is in
the paramagnetic phase and the entropy correctly shows a single
hump centered around zero magnetization. A second energy is in the
region of flips and the entropy shows three peaks, one centered at
zero and two symmetric ones centered at positive and negative
values of the magnetization. Finally, a third energy in the
ferromagnetic phase shows a single peak at a positive value of the
magnetization. It is likely that for this energy value we are in
presence of magnetization gaps, i.e. ergodicity breaking, since a
symmetric value of magnetization should be present at this energy,
but cannot be reached using microcanonical dynamics. Like for sc
lattices, we have checked whether the phase transition persists if
one increases the size of the base of the bcc lattice. With four
spins in the transversal layers, we get $K/J=-3/(2\pi)>-0.5$,
which predicts that magnetized states can be realized in the bcc
lattice even for large bases and for large aspect ratios. This is
confirmed in numerical simulations.

Finally, let us switch to {\it face centered cubic lattices}. In
the simplest realization of this lattice, there are four and five
spins in subsequent transversal layers (see Fig.~\ref{lattice}).
Numerical simulations show the same phenomenology as the one of
bcc lattices, for the smaller lengths. However, by looking at the
effective Hamiltonian for this lattice, we can predict that, like
for the sc lattice, magnetization does not persist for larger
bases. Indeed, each spin interacts with four neighbors inside a
transversal layer and, thus, the antiferromagnetic coupling
constant $K=-8\sqrt{2}\varepsilon$. The volume per spin is
$v_0=a^3/4$ and therefore $J=16\pi\varepsilon/3$ for large bases and
large aspect ratios. Consequently, $K/J=-3/(\sqrt{2}\pi)< -0.5$,
which excludes the presence of spontaneous magnetization.

As the length of the sample increases the system becomes more and
more one-dimensional. One might therefore doubt about the existence of
spontaneous magnetization for large aspect ratios, because dipolar
force is short-range in one dimension. However, it is well known that,
while one dimensional systems do not spontaneously magnetize, they nevertheless have a diverging
correlation length at small temperatures $T$,
$\ell=-a/\ln(\tanh(g/T))$, where $g$ is the short-range coupling
constant. We would like to give an estimate
of this correlation length to compare it with the sample lengths
that we use. First of all, one can get an estimate of the
temperature by treating canonically the single spin in interaction
with the thermal bath of all other spins. In the mean-field
approximation~\cite{abragam}, $m_z =\tanh\left[\mu H/T\right] $
where $H=(K+J) m_z/\mu$ is assumed to be constant over the whole
lattice. In our simulations for bcc lattices, the minimal
magnetization for which the ferromagnetic state survives is in the
range $ m_z\simeq 0.65$. Using the mean-field formula above, we
get the approximate value of the temperature of the system:
$T\simeq 2\varepsilon$. The corresponding short-range coupling
constant is $g=(K+J)/2\approx\varepsilon$, from which we get the
value of the correlation length  $\ell\approx a$. This value is
much smaller than the typical length of the sample, and then we
can conclude that the magnetization that we observe is not of
short-range origin.

In conclusion, we have shown clear evidences of the presence of
spontaneous magnetization in finite needle-like samples, within
the microcanonical ensemble. We believe that the origin of this
effect is in the long-range character of Hamiltonian~(\ref{11}), which
we were able to map onto an effective one-dimensional Ising model
with competing antiferromagnetic short-range  and ferromagnetic
mean-field couplings. The presence of jumps in magnetization as
energy is varied, the coexistence of paramagnetic and
ferromagnetic phases in some energy ranges, and the appearance of
flips of magnetization are all indications that the phase
transition we observe is of the first-order. We have simulated
three different kinds of cubic lattices and all of them show
spontaneous magnetization in some parameter ranges. Flips are
found only for bcc and fcc lattices.

Magnetization flips which have features of telegraph noise have
been observed experimentally for short-range ferromagnets~\cite{WernsdorferBarabra}.
Since rare-earth compounds ~\cite{ros,Corruccini} and cobalt
nanoparticle assemblies \cite{ref} are dominated by long-range dipolar interactions,
it could be extremely interesting to check experimentally the presence of
magnetization flips in isolated purely dipolar samples.

Our numerical simulations also show that, in the ferromagnetic phase, gaps in the accessible
values of magnetization appear. This is an indication of ergodicity breaking \cite{mukruf,celardo,bouchet},
one of the exotic properties of systems with long-range interactions \cite{rep} that could
also be checked in experiments performed in microcanonical conditions.

\begin{acknowledgments}
We thank B. Barbara, G. Celardo, D. Mukamel and P. Politi for helpful discussions. This
work has been partially supported by the joint grant EDC25019 from
CNRS (France) and SRNSF (Georgia),  by the grant ANR-10-CEXC-010-01
and by the SRNSF grant No 30/12. Numerical simulations were done at PSMN,
ENS-Lyon.
\end{acknowledgments}

\end{document}